\documentstyle[12pt]{article}

\newcommand{\be}{\begin{equation}}
\newcommand{\ee}{\end{equation}}
\newcommand{\bea}{\begin{eqnarray}}
\newcommand{\eea}{\end{eqnarray}}
\newcommand{\gsim}{ \mathop{}_{\textstyle \sim}^{\textstyle >} }
\newcommand{\lsim}{ \mathop{}_{\textstyle \sim}^{\textstyle <} }
\newcommand{\vev}[1]{ \left\langle {#1} \right\rangle }
\newcommand{\EV}{~\mbox{eV}} \newcommand{\kEV}{~\mbox{keV}}
\newcommand{\MEV}{~\mbox{MeV}} \newcommand{\GEV}{~\mbox{GeV}}
\newcommand{\TEV}{~\mbox{TeV}}

\def\al{\alpha}

\def\Frac#1#2{{\displaystyle\frac{#1}{#2}}}

\newcommand{\mod}{{\rm mod} \hspace{2mm}}
\newcommand{\fb}{\bar{f}}
\newcommand{\nb}{\bar{n}}
\newcommand{\hb}{\bar{h}}
\newcommand{\z}{{\bf Z}}
\newcommand{\half}{\frac{1}{2}}

\newcommand{\xb}{\bar{\xi}}

\begin{document}
\setlength{\baselineskip}{0.6cm}

\begin{titlepage}

\begin{flushright}
DESY-03-018\\
UT-03-06\\
\end{flushright}
 
\vskip 2cm
\begin{center}
 {\Large\bf Nonanomalous Discrete R-Symmetry
 \\
 and Light Gravitino}
 \vskip 1.2cm
 {\large Koichi Hamaguchi$^1$ and Nobuhito Maru$^2$}

 \vskip 0.4cm

 {\it $^1$ Deutsches Elektronen-Synchrotron DESY, D-22603, Hamburg, Germany}\\
 {\it $^2$ Department of Physics, University of Tokyo,
 Tokyo 113-0033, Japan}
\vskip 2cm
\abstract{ We discuss nonanomalous R-symmetry in the supersymmetric
grand unified theories.  In particular, we explore anomaly-free
solutions predicting the gravitino mass in the range of $10^{-3}\EV
\lsim m_{3/2} \lsim 1\TEV$ when the $\mu$-parameter is fixed to be
$\mu\simeq 1\TEV$.  In the minimal SU(5) GUT, we have shown that
$\mu\simeq 1\TEV$ is obtained only if the gravitino is ultralight with
mass $m_{3/2}\sim 10^{-3}\EV$. If extra fields ${\bf 5} \oplus {\bf
5^*}$ or ${\bf 10} \oplus {\bf 10^*}$ are introduced, many solutions
predicting $m_{3/2}\gsim 10^{-3}\EV$ are found. The R-parity is
violated due to the vacuum expectation value of the superpotential,
but it is controlled by the discrete R-symmetry. We find that the
R-parity violating couplings are naturally suppressed much below the
experimental bounds for some charge assignments. These charge
assignments predict light gravitino with masses of order ${\cal
O}(10^{-3}\EV)$--${\cal O}(1\MEV)$. These discrete R-symmetries can be
considered as solutions to the $\mu$-problem in low energy
supersymmetry breaking models such as the gauge mediation.}
\end{center}
\end{titlepage}

\section{Introduction}
A discrete R-symmetry $Z_{NR}$ often appears as a remnant of the
rotational symmetry of the compactified extra space in higher
dimensional supergravity or string theory \cite{Polchinski,IWY}.  This
discrete R-symmetry should be nonanomalous since this is a gauge
symmetry.  An R-symmetry plays a crucial role in the phenomenology of
supersymmetric (SUSY) theory.  First, it can suppress the cosmological
constant compared to the Planck scale.  Second, the SUSY-invariant
mass term (the $\mu$-term) of the Higgs chiral multiplet can be
forbidden so that the Higgs mass not be the Planck scale.  If an
R-symmetry breaking is related to SUSY breaking, the Higgs chiral
multiplet can obtain a mass of the order of the gravitino mass
$m_{3/2} \simeq 1\TEV$ by the Giudice-Masiero (GM) mechanism
\cite{GM}.  Third, an R-parity forbids the dimension-four baryon and
lepton number violating operators causing too rapid proton decay
\cite{P_decay,dim_5}.  These observations motivate us to ask whether we can find a
nonanomalous discrete R-symmetry with the above properties.  In a
paper by Kurosawa, Maru and Yanagida \cite{KMY}, nonanomalous discrete
R-symmetries in the minimal SUSY standard model (MSSM) and the SUSY
grand unified theory (GUT) were found under the situation that the GM
mechanism works.  These solutions can also forbid the dimension-five
baryon- and lepton-number violating operators.  Furthermore, extra
fields ${\bf 5} \oplus {\bf 5^*}$ with the mass of order $1\TEV$,
which can be testable in collider experiments, are predicted from the
anomaly cancellations in the GUT case.

In this paper, we consider nonanomalous R-symmetries {\em without} the
 GM mechanism.\footnote{We do not consider a possibility of anomaly
 cancellations by Green-Schwarz mechanism \cite{GS}.} In particular,
 we consider that the $\mu$-term is induced by the vacuum expectation
 value (VEV) of the superpotential $\vev{W}$, assuming that a
 fractional power of $\vev{W}$ is allowed in the superpotential. The
 fractional power makes it possible to obtain the correct size of the
 $\mu$-term even for gravitino mass smaller than the electroweak
 scale. We find that the $\mu\simeq 1\TEV$ is obtained only if the
 gravitino is extremely light as $m_{3/2}\simeq 10^{-3}\EV$ in the
 minimal SU(5) GUT.  If extra fields such as ${\bf 5} \oplus {\bf
 5^*}$ or ${\bf 10} \oplus {\bf 10^*}$ are added to the minimal SU(5)
 GUT, we find many charge assignments predicting $m_{3/2}\gsim
 10^{-3}\EV$. These charge assignments can be considered as solutions
 to the $\mu$-problem in low energy SUSY breaking models such as the
 gauge mediated SUSY breaking (GMSB)
 models~\cite{GMSB}.\footnote{Solutions to the $\mu$-problem in the
 GMSB models have been reported so far in Refs.~\cite{mu1,mu2,mu3,mu4}.
 Also interesting mechanisms for the $\mu$-problem in various
 mediation mechanisms of SUSY breaking are recently proposed
 \cite{KO,HNP}.}

In our framework, the R-parity is in general violated due to the
fractional powers of $\vev{W}$.  However, the R-parity violating
couplings are well controlled by the symmetry.  In fact, it turns out
that R-parity violation is small enough for some charge assignments.
It is further shown that the dimension-five baryon- and lepton-number
violating operators are naturally suppressed. In these charge
assignments the gravitino masses are predicted in the range
$m_{3/2}\sim {\cal O}(10^{-3}\EV)$--${\cal O}(1\MEV)$, and hence the
gravitino is the lightest supersymmetric particle (LSP).  Its lifetime
is much longer than the age of the universe, and the gravitino can be
the dominant component of the dark matter.

This paper is organized as follows.  In the next section, after
discussing the general constraint on the power of the superpotential,
we search for anomaly-free solutions of the minimal SU(5), and the
minimal SU(5) with ${\bf 5}\oplus{\bf 5^*}$ or ${\bf 10} \oplus {\bf
10^*}$.  In our analysis, constraints on the R-parity violating
operators from the proton decay and neutrino masses are taken into
account.  A brief discussion on the cosmology of the gravitino LSP
with R-parity violation is also given.  The last section contains a
summary of our paper.  In the appendix, a subtle issue between the
fractional power and the discrete symmetry is discussed.

\section{Discrete R-symmetry in GUT}

In this section, we consider the discrete R-symmetry in the GUT.
 Before discussing anomaly cancellations in detail, we give a general
 constraint on powers of the VEV of the
 superpotential in the next subsection.
 
\subsection{General constraint on the powers of $W$}

We assume that the $\mu$-term is generated from 
\bea
\label{m1}
W \simeq \left( \frac{\langle W \rangle}{M_P^3} \right)^y M_P H\bar{H}
\,, 
\eea
where $y$ is a non-negative number and $M_P = 2.4 \times 10^{18}$ GeV
 is the reduced Planck scale.  Here and hereafter, we omit
 coefficients of order unity. Then, the following conditions should be
 satisfied,
\bea
\label{mu1}
2 \al y + h + \bar{h} &=& 2\al   \quad(\mod N)\,, \\
\label{mu2}
h + \bar{h} &\ne& 2 \al   \quad(\mod N)\,, 
\eea
where $h, \bar{h}$ and $\alpha$ denote the R-charge of $H, \bar{H}$
 and the Grassmann coordinate $\theta$, respectively and they are all
 integers. (See Table \ref{GUT}.)  The second condition (\ref{mu2}) is
 necessary to forbid the Higgs mass term with Planck scale.  {}From
 Eq.~(\ref{m1}), we can predict the gravitino mass because
\bea
\mu \simeq \left( \frac{\langle W \rangle}{M_P^3} \right)^y M_P 
\simeq \left( \frac{m_{3/2}}{M_P} \right)^y M_P
\label{EQ-mu}
\,, 
\eea
and hence
\bea
\label{g32}
m_{3/2} \simeq \left( \frac{\mu}{M_P} \right)^{1/y} M_P 
\simeq 10^{18.4-15.4/y}\GEV\,, 
\eea
where $\mu \simeq \TEV$ is assumed throughout this paper. 
Eq.~(\ref{g32}) can be rewritten as 
\bea
y \simeq \frac{{\rm log}(M_P/\mu)}{{\rm log}(M_P/m_{3/2})}\,. 
\label{EQ-x}
\eea
{}From this equation, we can derive lower and upper bounds on $y$.  We
 assume
\begin{eqnarray}
  m_{3/2} \lsim \mu \ll M_P\,.
  \label{EQ-relations}
\end{eqnarray}
Thus, Eqs.(\ref{EQ-x}) and (\ref{EQ-relations}) lead to 
\begin{eqnarray}
  0 < y \lsim 1\,.
  \label{bound-on-y-1}
\end{eqnarray}
Here, one might wonder if the fractional power $y<1$ is incompatible
with the discrete symmetry. This issue is briefly discussed in the
Appendix. As can be seen from Eq.~(\ref{EQ-mu}), the fractional power
$y<1$ is crucial to obtain the correct size of $\mu$-term for a
gravitino mass smaller than the weak scale.

Besides the constraint in Eq.(\ref{bound-on-y-1}), there is a lower
bound on $y$ coming from the lower bound on the gravitino mass
$m_{3/2}$. In the GMSB model, when we fix the soft mass scale $m_{\rm
soft}$, the SUSY breaking $F$-term is bounded from below as $\sqrt{F}
\gsim {\cal O}(10\TEV)\times (m_{\rm soft}/100\GEV)$ in order to avoid
the negative mass squared for the scalar field in the messenger
sector~\cite{GMSB}.  This leads to a lower bound on the gravitino
mass, $m_{3/2}\gsim {\cal O}(0.01\EV)$. Even lighter gravitino mass
$m_{3/2}\sim {\cal O}(10^{-3}\EV)$ can be allowed in some SUSY
breaking models in higher dimensional
spacetime~\cite{SuperLightGrav}. For $m_{3/2}\simeq 10^{-3}\EV$, one
can see from Eq.~(\ref{EQ-x}) that $y$ can be as small as $y\simeq
1/2$.  {}From the above arguments, we impose the following constraints
on the parameter $y$:
\begin{eqnarray}
  \frac{1}{2} \le y \le 1\,.
\end{eqnarray}
%

\subsection{Anomaly cancellation}

\begin{table}[t!]
 \begin{center}
  \begin{tabular}{|c|rrrrrr|}
   \hline \raisebox{0ex}[12pt]{}
   \makebox[8mm]{} & 
   \makebox[5mm]{\hfill $T$} & \makebox[5mm]{\hfill $\bar{F}$} &
   \makebox[5mm]{\hfill $\bar{N}$} & \makebox[5mm]{\hfill $H$} &
   \makebox[5mm]{\hfill $\bar{H}$} & \makebox[5mm]{\hfill $\theta$} \\ 
   \hline
   $SU(5)$ & ${\bf 10}$ & ${\bf 5^*}$ & ${\bf 1}$ & ${\bf 5}$ & 
   ${\bf 5^*}$ & \\
   $Z_{NR}$ & $t$ & $\bar{f}$ & $\nb$ & $h$ & $\hb$ & $\al$ \\
   \hline
  \end{tabular}
  \caption{The matter content of GUT. $Z_{NR}$ charges of the fields
  denote those of the scalar components and are integers.  We take the
  R-charge of the Grassmann coordinate $\theta$ to be an integer
  $\alpha$.}
  \label{GUT}
 \end{center}
\end{table}

We are now at the position to discuss the anomaly cancellation.  Let
us first take the minimal SU(5) GUT.  Its matter content is described
in Table \ref{GUT}.  $Z_{NR}$ charge of the fields, which is taken to
be generation independent for simplicity, denotes those of the scalar
component and are integers.  Note that we take the R-charge of the
Grassmann coordinate $\theta$ to be an arbitrary integer $\alpha$.
The Yukawa couplings and the Majorana mass term are given
by\footnote{We have suppressed ${\cal O}$(1) coefficients of the
terms.}
\bea
\label{yukawaGUT}
W = TTH + T\bar{F}\bar{H} + \bar{F}\bar{N}H 
+ \frac{1}{2} M_m \bar{N}^2
\,, 
\eea
where $M_m$ is a Majorana mass for the right-handed neutrinos. 
For Eq.~(\ref{yukawaGUT}) to be allowed, 
 the corresponding R-charges have to satisfy 
\bea
\label{Yukawa-TTH}
&&2t + h = 2 \al   \quad(\mod N)\,, \\
\label{Yukawa-TFH}
&&t + \bar{f} + \bar{h} = 2 \al   \quad(\mod N)\,, \\
\label{Yukawa-FNH}
&&\fb + \nb + h = 2\al   \quad(\mod N)\,, \\
\label{RHN-mass}
&&2 \bar{n} = 2 \al   \quad(\mod N)\,. 
\eea
Anomaly cancellation conditions for $Z_{NR}[SU(3)_C]^2$ and 
$Z_{NR}[SU(2)_L]^2$ are \cite{IR} 
\bea
Z_{NR}[SU(3)_C]^2 &=& 
 \frac{3}{2} \{ 3(t - \al) + (\fb - \al) \} 
 + 3\al = \frac{N}{2}k   \quad(k \in \z)\,, \\
Z_{NR}[SU(2)_L]^2 &=& 
\frac{1}{2} \{ 9(t - \al) + 3(\fb - \al) \} 
+ \half \{ (h -\al) + (\hb - \al) \} 
+ 2 \al \nonumber \\
&=& \frac{N}{2}k' \quad(k' \in \z)\,. 
\eea
These conditions are simplified to 
\bea
\label{su3g}
h + \hb &=& 2\al   \quad(\mod \frac{N}{3})\,, \\
\label{su2g}
h + \hb &=& \al   \quad(\mod \frac{N}{2})\,. 
\eea
Eqs.~(\ref{su3g}) and (\ref{su2g}) lead to 
\bea
\label{rh}
h + \hb = 4 \al   \quad(\mod N)\,. 
\label{hhbar4al-mini}
\eea
Substituting this back into (\ref{su3g}) or (\ref{su2g}) results in
$6\al = 0~(\mod N)$, or equivalently
\bea
6\al = Nk   \quad(k \in \z)\,. 
\eea
Taking into account $0 < \alpha < N$, $0 < k < 6$ is obtained. 
If we take $k = 3$, 
 then $2\al = N$, and  
\bea
h + \hb = 4\al
= 2\al   \quad(\mod 2\al)\,, 
\eea
is derived and contradicts the condition (\ref{mu2}).  Thus, we find $k
=$ 1, 2, 4 and 5.  In other words,
$N$ is classified as, 
\bea
N = 6\al,~3\al,~3 \left( \frac{\al}{2} \right),~
6 \left( \frac{\al}{5} \right). 
\eea
First, $N = 6\al$ case is considered. 
{}From Eqs.~(\ref{mu1}) and (\ref{hhbar4al-mini}), 
\bea
2 \al y + 4\al = 2\al   \quad(\mod 6\al) \to y = -1 + 3n \quad(n \in \z)\,. 
\eea
This has no solution for $1/2 \le y \le 1$.  Second case is $N = 3\al$.
{}From Eqs.~(\ref{mu1}) and (\ref{hhbar4al-mini}), we obtain
\bea
2 \al y + 4 \al = 2 \al   \quad(\mod 3 \al) 
\to y = -1 + \frac{3}{2}n \quad(n \in \z)\,. 
\eea
This has a solution $y = 1/2$ for $n = 1$. Third case is $N = 3\left(
 \al/2\right) = 3 \al'~(\al = 2\al',\,\al' \in \z)$.  From
 Eqs.~(\ref{mu1}) and (\ref{hhbar4al-mini}), we obtain
\bea
&&2 \al y + 4\al = 2\al   \quad(\mod 3\al')\,, \\ 
&&\Leftrightarrow 4 \al' y + 8 \al' = 4 \al'   \quad(\mod 3\al') 
\to y = -1 + \frac{3}{4}n \quad(n \in \z)\,. 
\eea
This also has a solution $y = 1/2$ for $n = 2$. The last case is $N =
 6\left(\al/5 \right) = 6 \al'~(\al = 5\al',\,\al' \in \z)$.
 From Eqs.~(\ref{mu1}) and (\ref{hhbar4al-mini}), we obtain
\bea
&&2 \al y + 4\al = 2\al   \quad(\mod 6\al')\,, \\
&& \Leftrightarrow 10 \al' y + 20 \al' = 10 \al'   \quad(\mod 6\al') 
\to y = -1 + \frac{3}{5}n \quad(n \in \z)\,. 
\eea
Taking into account $1/2 \le y \le 1$, $5/2 < n \le 10/3$ is obtained.
 This has a solution $n=3$, but it does not correspond to the minimum
 solution for positive $y$.  The dominant $\mu$-term comes from the
 term with $y=1/5$ (i.e. $n=2$), and it would cause $\mu\gg 1\TEV$
 unless $m_{3/2}\ll 10^{-3}\EV$. Thus, this case has no solution.

Therefore, the correct size of the $\mu$-term can be obtained for
$y=1/2$ when $N = 3\alpha$ and $N = 3(\alpha/2)$, which predict an
extremely light gravitino with mass $m_{3/2}\simeq {\cal O}(10^{-3}\EV)$.

\subsubsection*{Baryon- and lepton-number violating operators}

We have shown that the correct size of $\mu$-term can be obtained for
ultralight gravitino $m_{3/2}\simeq 10^{-3}\EV$, by means of the
fractional power of the superpotential's VEV,
$\vev{W}^{1/2}$. However, if we allow general interaction terms
including the fractional power of $\vev{W}$, there appear baryon- and
lepton-number violating operators as well. Therefore we have next to
consider constraints on these operators.

Let us first consider the following superpotential
\begin{eqnarray}
  W \simeq \left(\frac{\vev{W}}{M_P^3}\right)^z M_P\bar{F} H\,,
\end{eqnarray}
which includes the so-called bilinear R-parity violation, $W =
\widehat{\mu}_i L_i H_u$.  The coupling is given by
\begin{eqnarray}
  \widehat{\mu}\simeq \left(\frac{m_{3/2}}{M_P}\right)^z M_P\,.
  \label{EQ-bilinear}
\end{eqnarray}
The bilinear R-parity violation can generate the neutrino
mass~\cite{Hall-Suzuki} which explains the atmospheric neutrino
oscillation if $\widehat{\mu}/\mu \sim {\cal
O}(10^{-4}$--$10^{-7})$~\cite{Private,bilinear}. In other words,
$\widehat{\mu}$ should be smaller than ${\cal
O}(10^{-1}\GEV$--$10^{-4}\GEV)$ in order to avoid a too large neutrino
mass.

The power $z$ is determined by the symmetry:
\begin{eqnarray}
  && 2\alpha z + \bar{f} + h = 2\alpha   \quad(\mod N)\,,
  \nonumber\\
  &\to& 2\alpha z = \bar{n}   \quad(\mod N)\,,
  \nonumber\\
  &\to& z = \frac{\bar{n} + rN}{2\alpha}\quad (r\in {\bf Z})\,,
  \label{EQ-z-for-FH}
\end{eqnarray}
where we have used Eq.~(\ref{Yukawa-FNH}). {}From
Eq.~(\ref{RHN-mass}), the charge of the right-handed neutrino should
be either $\bar{n}=\alpha$ or $\bar{n}=\alpha + N /2~(\mod N)$. If
$\bar{n}=\alpha$, the minimum non-negative $z$ is given by $z = 1/2$
since $N > \alpha$. In this case, the bilinear coupling is given by
\begin{eqnarray}
  \widehat{\mu}\simeq \left(\frac{m_{3/2}}{M_P}\right)^{1/2}M_P
  \simeq 10^3\GEV\left(\frac{m_{3/2}}{10^{-3}\EV}\right)^{1/2}
  \,.
\end{eqnarray}
This generates too large neutrino mass and hence is excluded.  Thus,
the charge of the right-handed neutrino should be $\bar{n}=\alpha + N
/2~(\mod N)$. This also means that the case of $N = $ odd is
excluded. If $N = $ even and $\bar{n}=\alpha + N /2~(\mod N)$,
Eq.~(\ref{EQ-z-for-FH}) gives rise to
\begin{eqnarray}
  z = 
  \left\{
  \begin{array}{ccl}
    \Frac{5 + 6r}{4}& {\rm for}& N = 3\alpha\,,
    \\
    &&
    \\
    \Frac{7 + 6r}{8}& {\rm for}& N = \Frac{3\al}{2}\,.
    \end{array}
  \right.
\end{eqnarray}
The minimum non-negative $z$ are given by $z = 5/4$ ($r = 0$) and $z =
1/8$ ($r = -1$), respectively. Therefore, the bilinear R-parity
violating coupling is given by
\begin{eqnarray}
  \widehat{\mu}\simeq
    \left\{
    \begin{array}{ccl}
      \left(\Frac{m_{3/2}}{M_P}\right)^{5/4}M_P
      \,\simeq\, 
      3\times 10^{-20}\GEV
      \left(\Frac{m_{3/2}}{10^{-3}\EV}\right)^{5/4}
      & {\rm for}& N = 3\alpha\,,
      \\
      &&
      \\
      \left(\Frac{m_{3/2}}{M_P}\right)^{1/8}M_P
      \,\simeq\, 
      4\times 10^{14}\GEV
      \left(\Frac{m_{3/2}}{10^{-3}\EV}\right)^{1/8}
      & {\rm for}& N = \Frac{3\alpha}{2}\,.
      \end{array}
    \right.
\end{eqnarray}
Thus, the $N = 3(\al/2)$ case is clearly excluded. On the other hand,
for $N = 3\al$ the contribution to the neutrino mass from the R-parity
violation is extremely small. Therefore, the dominant contribution to
the neutrino masses are understood to be generated by the standard
seesaw mechanism~\cite{seesaw}.

Next we consider the trilinear R-parity violation caused by the
following superpotential
\begin{eqnarray}
  W \simeq \left(\frac{\vev{W}}{M_P^3}\right)^{z'}T\bar{F}\bar{F}
  \simeq \left(\frac{m_{3/2}}{M_P}\right)^{z'}T\bar{F}\bar{F}
  \,\equiv\,
  \lambda_{\rm eff}T\bar{F}\bar{F}\,,
\end{eqnarray}
which includes both of the baryon- and lepton-number violating
operators, $\lambda_{ijk} L_i L_j E^c_k$, $\lambda'_{ijk} L_i Q_j
D^c_k$ and $\lambda_{ijk}''U^c_i D^c_j D^c_k$. The power $z'$ is
determined by
\begin{eqnarray}
  && 2\alpha z' + t + 2\bar{f} = 2\alpha\quad(\mod N)\,,
  \nonumber \\
  &\to& 2\alpha z' = 2\alpha + \bar{n}\quad(\mod N)\,,
  \label{EQ-zprime}
\end{eqnarray}
where we have used Eqs.~(\ref{Yukawa-TFH}), (\ref{Yukawa-FNH}) and $h
+ \bar{h} = 4\alpha$. For $N = 3\al$ ($=$ even) and $\bar{n}=\alpha +
N /2~(\mod N)$, $z'$ is given by
\begin{eqnarray}
  z' = \frac{9 + 6r'}{4}\quad (r'\in {\bf Z})
  \,.
\end{eqnarray}
The minimum non-negative $z'$ is given by $z'=3/4$ ($r'=-1$) and the
effective coupling becomes
\begin{eqnarray}
  \lambda_{\rm eff}\simeq 10^{-23}\left(\frac{m_{3/2}}{10^{-3}\EV}\right)^{3/4}
  \,,
\end{eqnarray}
which easily satisfy the constraint from the proton decay,
$\lambda'_{11j} \lambda''_{11j} \lsim 2\times 10^{-27} (m_{\rm
soft}/100\GEV)^2$ $(j=2,3)$~\cite{lambda-lambda}.

Let us also discuss the dimension five operator $W \simeq (1/M_{\rm
eff})TTT\bar{F}$, which causes the proton decay~\cite{dim_5}.  In
order to suppress the proton decay rate below the experimental bound,
the effective mass scale should be $M_{\rm eff} > {\cal
O}(10^{25}\GEV)$~\cite{Meff-bound}, i.e., much larger than
$M_P$. Notice that the R-parity cannot forbid this operator. In our
framework, the operator is given by
\begin{eqnarray}
  W \simeq \frac{1}{M_P}\left(\frac{\vev{W}}{M_P^3}\right)^{z''}TTT\bar{F}
  \simeq \frac{1}{M_P}\left(\frac{m_{3/2}}{M_P}\right)^{z''}TTT\bar{F}
  \,\equiv\,
  \frac{1}{M_{\rm eff}}TTT\bar{F}\,,
  \label{EQ-Meff}
\end{eqnarray}
and $z''$ should satisfy
\begin{eqnarray}
  && 2\alpha z'' + 3t + \bar{f} = 2\alpha\quad(\mod N)\,,
  \nonumber \\
  &\to& 2\alpha z'' = 2\alpha\quad(\mod N)\,,
  \nonumber \\
  &\to& z'' = \frac{2\alpha + r''N}{2\alpha}
  = \frac{2 + 3r''}{2}\quad(r''\in {\bf Z})\,,
\end{eqnarray}
where we have used Eqs.~(\ref{Yukawa-TTH}), (\ref{Yukawa-TFH}), $h +
\bar{h} = 4\alpha$, and $N = 3\al$. The minimum non-negative $z''$ is
given by $z''=1$ ($r''=0$), and hence the effective mass scale $M_{\rm
eff} \sim M_P(M_P/m_{3/2})\sim 10^{48}\GEV(10^{-3}\EV/m_{3/2})$ is much
above the experimental bound.  Therefore, the discrete R-symmetry
naturally suppresses the dimension five proton decay operator.

\subsubsection*{explicit $Z_{NR}$ charge assignments}

\begin{table}[t!]
 \begin{center}
  \begin{tabular}{|c|cccccc|}
   \hline \raisebox{0ex}[11pt]{}
   & $t$ & $\fb$ & $\nb$ & $h$ & $\hb$ & $\theta$ \\
   \hline
   $P$ & $1$ & $1$ & $1$ & $0$ & $0$ & $\al$ \\
   $V$ & $1$ & $-3$ & $5$ & $-2$ & $2$ & \\
   $A$ & $0$ & $-1$ & $1$ & $0$ & $1$ & \\
   \hline
  \end{tabular}
 \end{center}
 \caption{The charges of the generators $P, V, A$.}
 \label{gen}
\end{table}

Before closing this subsection, we comment on the explicit form of
$Z_{NR}$.  Nonanomalous R-symmetries are represented by $Z_{NR} =
P^\alpha V^\beta A^\gamma~(\alpha, \beta, \gamma \in \z)$ \cite{IR}
where the generators $P, V$ and $A$ are summarized in Table \ref{gen}.
One can easily check that the charge assignments in Table \ref{gen}
satisfy Yukawa conditions
(\ref{Yukawa-TTH})-(\ref{Yukawa-FNH}).\footnote{By imposing 3
conditions (\ref{Yukawa-TTH})-(\ref{Yukawa-FNH}) on the 6 parameters
$t$, $\fb$, $\nb$, $h$, $\hb$ and $\theta$, the charge assignments can
be represented in terms of 3 parameters $\al$, $\beta$ and $\gamma$.}
For $N = 3\al$ ($=$ even) and $\bar{n}=\alpha + N /2~(\mod N)$, we
obtain
\bea
\label{gen01}
\nb = \al + 5\beta + \gamma = \al +\frac{N}{2}\quad({\rm mod}~N) 
&\Leftrightarrow& 5\beta + \gamma = \frac{N}{2} = \frac{3}{2}\al\quad({\rm mod}~N)
\,,
\eea
while Eq.~(\ref{hhbar4al-mini}) leads to
\bea
\label{gen02}
-2\beta + (2\beta + \gamma) = \gamma = 4\al \quad({\rm mod}~N)\,.
\eea
From Eqs.~(\ref{gen01}) and (\ref{gen02}), we obtain
\bea
Z_{3\al R} &=& (PA^4)^\al V^\beta\,, \\
5\beta &=& -\frac{5}{2}\al \quad({\rm mod}~N=3\al)
\nonumber\\
&=& 
-\frac{5}{2}\al + 3\al m\quad(m\in {\bf Z})\,.
\eea
Therefore,
\bea
\al : \beta : N = 10 : (-5+6m) : 30\,,
\eea
and the explicit forms of the $Z_{NR}$ are given by
\bea
Z_{30R}
&=&
(PA^4)^{10} V^{6m-5} \quad(m = 1,2,3,4)\,,
\label{EQ-Z30R}
\\
Z_{6R}
&=&
(PA^4)^{2} V^5\,.
\label{EQ-Z6R}
\eea
%

\subsection{Introducing ${\bf 5} \oplus {\bf 5^*}$}

In this subsection and the next, we consider the possibility that
 $10^{-3}\EV \lsim m_{3/2} \lsim 1\TEV$ is predicted by introducing a
 pair of ${\bf 5} \oplus {\bf 5^*}$ or ${\bf 10} \oplus {\bf 10^*}$.
 If we introduce a pair of ${\bf 5} \oplus {\bf 5^*}$ with $Z_{NR}$
 charges $\xi$ and $\xb$, respectively, the anomaly cancellation
 conditions are modified as
\bea
Z_{NR}[SU(3)_C]^2 &=& 
 \frac{3}{2} \{3(t - \al) + (\fb - \al)\} 
 + \half \{(\xi - \al) + (\xb - \al)\} + 3\al 
 = \frac{N}{2}k \quad(k \in {\bf Z}) \nonumber \\
 &\Leftrightarrow& 3(h + \hb) - (\xi + \xb) = 4\al   \quad(\mod N)\,, \\
 Z_{NR}[SU(2)_L]^2 &=& 
 \frac{1}{2} \{9(t - \al) + 3(\fb - \al)\} + \half (h + \hb - 2\al) 
 \nonumber \\
 &&+ \half \{(\xi - \al) + (\xb - \al)\}
 + 2 \al = \frac{N}{2}k' \quad(k' \in \z)  \nonumber \\
 &\Leftrightarrow& 2(h + \hb) = \xi + \xb   \quad(\mod N)\,. 
 \eea
{}From these conditions, 
\bea
\label{su355}
&&h +\hb = 4\al   \quad(\mod N)\,, 
\label{EQ-hhbar-55}
\\
&&\xi + \xb = 8\al   \quad(\mod N)\,,
\label{EQ-xixi-55}
\eea
are obtained. 
At this stage, $N$ is undetermined. 
In order to fix $N$, let us take into account 
 the mixed gravitational anomaly cancellation~\cite{IR}, 
\bea
Z_{NR}[{\rm gravity}]^2 
&=& 30 (t - \al) + 15 (\fb - \al) + 3 (\nb - \al) 
+ 2 (h - \al) + 2 (\hb - \al) \nonumber \\ 
&& + 5 (\xi - \al) + 5 (\xb - \al) + (8 + 3 + 1) \al - 21\al~ \nonumber\\
&=& -23\al   \quad(\mod N~{\rm or}~N/2)\,. 
\eea
Mod $N~{\rm or}~N/2$ depends on whether $N$ is odd or even. In the
following, we consider whether the mixed anomaly can be canceled {\em
without introducing singlets}.

If $N$ is odd,
\bea
N = \frac{23}{k} \al \quad(k = 1,2,\cdots,22)\,. 
\eea
Eq.~(\ref{g32}) tells us 
\bea
m_{3/2} \simeq 10^{18.4 - \frac{15.4}{-1 + \frac{23}{2k}n}}
\GEV  \quad(n \in \z)\,,
\eea
because 
\bea
2 \al y + 4\al = 2\al   \quad(\mod 23\al/k) \to y = -1 + \frac{23}{2k}n\,. 
\eea
Taking into account that $1/2 \le y \le 1$ and $y$ is a minimum of
 non-negative value, we obtain
\bea
&&k= 6, 7 \quad({\rm for}~n=1)\,, \\
&&k= 12, 13, 14, 15 \quad({\rm for}~n=2)\,. 
\eea
The gravitino mass is summarized in Table \ref{table55-Pre}.

\begin{table}[t!]
 \begin{center}
  \begin{tabular}{|c|c|c|c|c|c|}
   \hline 
   $N:$~odd & & & & & \\
   \hline
   $(n, k)$ & $m_{3/2}$ & $(n, k)$ & $m_{3/2}$ & $(n, k)$ & 
   $m_{3/2}$ \\
   \hline
   (1,6) & 40~GeV & (2, 12) & 40~GeV & (2, 14) & 2.9~keV \\
   (1,7) & 2.9~keV & (2, 13) & 24~MeV & (2, 15) & 0.035~eV \\
   \hline
   $N:$~even & & & & & \\
   \hline
   (1, 12) & 40~GeV & (2, 24) & 40~GeV & (2, 28) & 2.9~keV \\
   (1, 13) & 24~MeV & (2, 25) & 1.2~GeV & (2, 29) & 14~eV \\
   (1, 14) & 2.9~keV & (2, 26) & 24~MeV & (2, 30) & 0.035~eV \\
   (1, 15) & 0.035~eV & (2, 27) & 330~keV &  &  \\
   \hline
  \end{tabular}
 \end{center}
 \caption{The gravitino mass for GUT with $5 \oplus 5^*$.}
 \label{table55-Pre}
\end{table}

If $N$ is even, 
\bea
&&
N = \frac{46}{k}\alpha \quad(k=1, 2,\cdots , 45)\,, 
\label{EQ-55even-N}
\\
&&
2 \al y + 4\al = 2\al   \quad(\mod 46\al/k) \to y = -1 + \frac{23}{k}n\,,
\eea
and Eq.~(\ref{g32}) leads to
\bea
m_{3/2} \simeq 10^{18.4 - \frac{15.4}{-1 + \frac{23}{k}n}}\GEV \quad(n \in \z)\,.
\eea
Taking into account that $1/2 \le y \le 1$ and $y$ is a minimum of
 non-negative value, we obtain
\bea
&&k= 12 \sim 15 \quad({\rm for}~n=1)\,, \\
&&k= 24 \sim 30 \quad({\rm for}~n=2)\,. 
\eea
The gravitino mass in this case is also summarized in Table \ref{table55-Pre}.

Now let us turn to discuss the baryon- and lepton-number violating
operators. First, the bilinear R-parity violating coupling
$\widehat{\mu}$ is given by Eq.~(\ref{EQ-bilinear}), with a fractional
power $z$ in Eq.~(\ref{EQ-z-for-FH}). One can show in the same way as
before that the case of $\bar{n} = \al$ is excluded since it results
in $z = 1/2$, which induces too large neutrino masses.  If $N = $ even
and $\bar{n}=\alpha + N /2~(\mod N)$, Eq.~(\ref{EQ-z-for-FH}) and
Eq.~(\ref{EQ-55even-N}) give rise to
\begin{eqnarray}
  z = \frac{k + 23 + 46 r}{2 k}\quad (r\in {\bf Z})\,.
\end{eqnarray}
The minimum non-negative $z$ is given by $r = 0$ for $k = 12$--$15$,
and $r = -1$ for $k = 24$--$30$. The resultant bilinear couplings are
given by $\widehat{\mu}\simeq (8\times 10^{-7}$--$7\times
10^{-19})\GEV$ for $k = 12$--$15$, and $\widehat{\mu}\simeq
(10^{15}$--$10^{18})\GEV$ for $k = 24$--$30$. Hence, $k = 24$--$30$
cases are excluded. Namely, the remaining charge assignments are $N =
(46/k)\al =$ even, $\bar{n}=\alpha + N /2~(\mod N)$, and $k = 12$--$15$.

Next we consider the trilinear R-parity violating coupling 
$W\sim \lambda_{\rm eff}T\bar{F}\bar{F}$. 
The effective coupling $\lambda_{\rm eff}$ is
given by $\lambda_{\rm eff}\simeq (m_{3/2}/M_P)^{z'}$. The power $z'$
is again determined by Eq.~(\ref{EQ-zprime}), $2\alpha z' = 2\alpha +
\bar{n}~(\mod N)$. (Notice that $h + \bar{h} = 4\alpha$ is satisfied
also in the present case. See Eq.~(\ref{EQ-hhbar-55}).) Therefore, for
$N = (46/k)\al =$ even and $\bar{n}=\alpha + N /2~(\mod N)$, $z'$ is
given by
\begin{eqnarray}
  z' = \frac{3k + 23 + 46r'}{2k}\quad (r'\in {\bf Z})\,.
\end{eqnarray}
The minimum non-negative $z'$ is given by $r'=0$ for $k = 12$--$15$,
and the trilinear couplings are $\lambda_{\rm eff}\simeq 8\times
10^{-10}$ ($k=12$), $5\times 10^{-13}$ ($k = 13$), $6\times 10^{-17}$
($k = 14$), and $7\times 10^{-22}$ ($k = 15$).  Thus, the cases of $k
= 12,13$ are excluded by the constraint from the proton decay,
$\lambda'_{11j}\lambda''_{11j} \lsim 2\times 10^{-27}\,(m_{\rm
soft}/100\GEV)^2~(j=2,3)$~\cite{lambda-lambda}.

The R-parity violating couplings are listed in Table.~\ref{table55}
for the remaining charge assignments $k = 14,15$ together with the
gravitino masses.  We find that the neutrino mass induced by these
R-parity violations is too small to explain the mass scale observed in
the atmospheric and solar neutrino oscillation experiments, and hence
the dominant contribution to the neutrino masses should be generated
by the seesaw mechanism.

\begin{table}[t!]
  \begin{center}
    \begin{tabular}{|c|c|c|c|}
      \hline
      $k$ & $m_{3/2}$ & $\widehat{\mu}$ & $\lambda_{\rm eff}$
      \\
      \hline
      14 & 2.9 keV & $6\times 10^{-14}\GEV$ & $6\times 10^{-17}$
      \\
      15 & 0.035 eV & $7\times 10^{-19}\GEV$ & $7\times 10^{-22}$
      \\
      \hline
    \end{tabular}
  \end{center}
  \caption{The gravitino mass for GUT with $5 \oplus 5^*$ and R-parity
  violating couplings $\widehat{\mu}$ and $\lambda_{\rm eff}$.  $N=$
  even and $\bar{n}=\alpha + N /2~(\mod N)$. Only the cases which
  predict sufficiently small R-parity violation are listed.  }
  \label{table55}
\end{table}

As for the dimension five operator $W \simeq (1/M_{\rm
eff})TTT\bar{F}$, the effective mass scale is given by
Eq.~(\ref{EQ-Meff}), $M_{\rm eff}\simeq M_P(M_P/m_{3/2})^{z''}$. It is
easy to show that the power is again given by $z''=1$ for $N =
(46/k)\al =$ even, $\bar{n}=\alpha + N /2~(\mod N)$ and $k =
14,15$. Thus, the effective mass scale $M_{\rm eff} \sim
M_P(M_P/m_{3/2})\sim 10^{39}\GEV(1\kEV/m_{3/2})$ is much above the
experimental bound and the proton decay via the dimension five
operator is naturally suppressed by the discrete R-symmetry also in
these charge assignments.

Finally, we comment on the explicit form of $Z_{NR} = P^\alpha V^\beta
A^\gamma~(\alpha, \beta, \gamma \in \z)$. (See Table \ref{gen}).  We
only consider the cases of $N =$ even, $\bar{n}=\alpha + N/2~(\mod N)$
and $k = 14,15$, since other cases are excluded by the constraints on
R-parity violation, as we have shown. Thus,
\bea
\label{gen1}
\nb = \al + 5\beta + \gamma = \al +\frac{N}{2}\quad({\rm mod}~N) 
&\Leftrightarrow& 5\beta + \gamma = \frac{N}{2} \quad({\rm mod}~N)\,. 
\eea
Eq.~(\ref{su355}) leads to 
\bea
\label{gen2}
-2\beta + (2\beta + \gamma) = \gamma = 4\al \quad({\rm mod}~N) 
\,.
\eea
From Eqs.~(\ref{gen1}) and (\ref{gen2}), 
\bea
\label{gen3}
5\beta + 4\al = \frac{N}{2} \quad({\rm mod}~N)\,,
\eea
is obtained. Thus, we have
\bea
Z_{\frac{46}{k}\al R} &=& (PA^4)^\al V^\beta\,, \\
5\beta + 4\al &=& \frac{23}{k}\al~(\mod \frac{46}{k}\al) 
= \frac{23}{k}\al + \frac{46}{k}\al m~(m \in {\bf Z})\,.
\eea
Therefore,
\bea
\al : \beta : N = 5k : (46m+23-4k) : 230\,,
\eea
and we obtain
\bea
Z_{230R}
&=&
(PA^4)^{70} V^{46m-33} \quad(k=14\,,\,\, m = 1,2,4,5)\,,
\nonumber\\
&&(PA^4)^{75} V^{46m-37}  \quad(k=15\,,\,\, m = 1,3,4,5)\,.
\\
Z_{46R}
&=&
(PA^4)^{14} V^{21}  \quad(k=14)\,,
\quad(PA^4)^{15} V^{11}  \quad(k=15)\,.
\eea
%

\subsection{Introducing ${\bf 10} \oplus {\bf 10^*}$}

Next, we consider the case with $\xi({\bf 10}) \oplus \xb({\bf 10^*})$. 
The anomaly cancellation conditions are modified as 
\bea
&Z_{NR}[SU(3)_C]^2:& 
 -\frac{3}{2} (h + \hb) + 3\al 
 + \frac{3}{2} (\xi + \xb) + \frac{3}{2}Nk" - 3\al = \frac{N}{2}k\,, \\
&Z_{NR}[SU(2)_L]^2:& 
- (h + \hb) + \al + \frac{3}{2} (\xi + \xb) - 3\al = \frac{N}{2}k'\,. 
\eea
These are simplified to 
\bea
3(h + \hb) &=& 3(\xi + \xb)   \quad(\mod N)\,, \\
2(h + \hb) &=& 3(\xi + \xb) - 4\al   \quad(\mod N)\,, 
\eea
and then 
\bea
&&h + \hb = 4\al   \quad(\mod N)\,, 
\label{EQ-hhbar-1010}\\
&&3(\xi + \xb) = 12\al   \quad(\mod N)\,, 
\label{EQ-1010-xixi}
\eea
are obtained. 
In order to fix $N$, let us take into account 
 the mixed gravitational anomaly cancellation, 
\bea
Z_{NR}[{\rm gravity}]^2 
&=& 30 (t - \al) + 15 (\fb - \al) +3 (\nb - \al) 
+ 2 (h - \al) + 2 (\hb - \al) + 10 (\xi - \al) \nonumber \\
&& + 10 (\xb - \al) + (8 + 3 + 1) \al - 21\al~
\nonumber\\
&=& (\xi + \xb) - 37\al   \quad(\mod N~{\rm or}~N/2)\,. 
\label{EQ-1010-grav}
\eea

In the same manner as the $\xi({\bf 5}) \oplus \xb({\bf 5^*})$ case,
one can show that the $\bar{n} = \alpha$ case is excluded because it
would generate too large neutrino mass from the bilinear R-parity
violation. Therefore, we consider the case of $N = $ even and
$\bar{n}=\alpha + N/2$. Then, Eqs.~(\ref{EQ-1010-xixi}) and
(\ref{EQ-1010-grav}) lead to
\bea
&&3Z_{NR}[{\rm gravity}]^2 = -99\al   \quad(\mod N/2)\,, \\
&&\to N = \frac{198}{k}\al \quad(k=1, 2,\cdots , 197)\,.
\label{EQ-N-1010}
\eea
{}From Eq.~(\ref{mu1}), we obtain 
\bea
y = -1 + \frac{99}{k}n\quad(n \in {\bf Z})\,. 
\eea
Taking into account that $1/2 \le y \le 1$ and $y$ is a minimum of
 positive values,
\bea
\label{ten1}
&&k = 50 \sim 66 \quad({\rm for}~n=1)\,, \\
\label{ten2}
&&k = 100 \sim 132 \quad({\rm for}~n=2)\,,
\eea
are obtained. 

Some of these charge assignments cause too large baryon- or
lepton-number violation via the R-parity violation. The orders of
magnitudes of bilinear ($\widehat{\mu}$) and trilinear ($\lambda_{\rm
eff}$) couplings can be estimated in the same way as ${\bf 5} \oplus
{\bf 5^*}$ case. In Table.~\ref{table-1010}, we show the gravitino
mass and these couplings for the cases in which the R-parity
violations are below the experimental bounds ($k = 57$--$66$).  In
these cases, the dimension five operator $W \simeq (1/M_{\rm
eff})TTT\bar{F}$ are naturally suppressed as $M_{\rm eff} \gg
10^{36}\GEV$ like the ${\bf 5} \oplus {\bf 5^*}$ case.

\begin{table}[t!]
  \begin{center}
    \begin{tabular}{|c|c|c|c|}
      \hline
      $k$ & $m_{3/2}$ & $\widehat{\mu}$ & $\lambda_{\rm eff}$
      \\
      \hline
      57 & 3.2 MeV & $7\times 10^{-11}\GEV$ & $7\times 10^{-14}$
      \\
      58 & 420 keV & $9\times 10^{-12}\GEV$ & $9\times 10^{-15}$
      \\
      59 & 49 keV & $1\times 10^{-12}\GEV$ & $1\times 10^{-15}$
      \\
      60 & 5.2 keV & $1\times 10^{-13}\GEV$ & $1\times 10^{-16}$
      \\
      61 & 490 eV & $1\times 10^{-14}\GEV$ & $1\times 10^{-17}$
      \\
      62 & 41 eV & $8\times 10^{-16}\GEV$ & $8\times 10^{-19}$
      \\
      63 & 2.9 eV & $6\times 10^{-17}\GEV$ & $6\times 10^{-20}$
      \\
      64 & 0.18 eV & $4\times 10^{-18}\GEV$ & $4\times 10^{-21}$
      \\
      65 & 0.95$\times 10^{-2}$ eV & $2\times 10^{-19}\GEV$ & $2\times 10^{-22}$
      \\
      66 & 0.4$\times 10^{-3}$ eV & $9\times 10^{-21}\GEV$ & $9\times 10^{-24}$
      \\
      \hline
    \end{tabular}
  \end{center}
  \caption{The gravitino mass for GUT with $10 \oplus 10^*$ and
  R-parity violating couplings $\widehat{\mu}$ and $\lambda_{\rm
  eff}$.  $N=$ even and $\bar{n}=\alpha + N /2~(\mod N)$. Only the
  cases which predict sufficiently small R-parity violation are
  listed.  }
  \label{table-1010}
\end{table}

Finally, we comment on the concrete form of $Z_{NR}$.  {}From
Eqs.~(\ref{EQ-hhbar-1010}), (\ref{EQ-N-1010}) and $\bar{n}=\al + N/2$, we obtain
\bea
Z_{\frac{198}{k}\al R} &=& (PA^4)^\al V^\beta\,, \\
5\beta + 4\al &=& \frac{99}{k}\al~(\mod N = \frac{198}{k}\al)
= \frac{99}{k}\al + \frac{198}{k}\al m~(m \in {\bf Z})\,,
\eea
and hence
\bea
\al : \beta : N = 5k : (198m+99-4k) : 990\,.
\eea
For $k = 66$, it reduces to $Z_{30R}$ or $Z_{6R}$ given in
Eqs.~(\ref{EQ-Z30R}) and (\ref{EQ-Z6R}). For $k = 57$--$65$, we obtain
\bea
Z_{110R} &=& (PA^4)^{35} V^{22m-17} \quad(k=63)\,,\\
Z_{330R} &=& (PA^4)^{95} V^{66m-43} \quad(k =57)\,,
\quad(PA^4)^{100} V^{66m - 47} \quad(k = 60)\,,\\
Z_{990R} &=& (PA^4)^{5k} V^{198m+99-4k} \quad({\rm other}~k)\,,
\eea
where $m = 1,2,3,4,5$. Among these cases, for $198m+99 - 4k = 5m'~(m'\in
{\bf Z})$, they reduce to $Z_{22R}$, $Z_{66R}$, and $Z_{198R}$.

\subsection{Gravitino LSP with R-parity violation}

As can be seen from Table.~\ref{table55} and \ref{table-1010}, the
 predicted masses of the gravitino are ${\cal O}(10^{-3}\EV$--$1\MEV)$.
 Therefore, the gravitino is the LSP.  Here, let us comment on the
 cosmology of this gravitino LSP.  Gravitino LSP dark matter without
 R-parity was investigated in Ref.~\cite{Taka-Yama} under the
 assumption that the R-parity violation is the dominant contribution
 to the neutrino masses. According to them, we have found that the
 lifetime of the gravitino is much longer than the age of the
 universe,\footnote{The lifetime of the LSP gravitino could be shorter
 than the age of the universe~\cite{Moreau-Chemtob} if there is a
 large trilinear coupling $\lambda\sim {\cal O}(0.1$--$1)$ close to
 the experimental bound and if the gravitino mass is relatively large,
 $m_{3/2} \gsim {\cal O}(1\GEV)$. However, this is not the case in our
 scenario.} since the decay rate is suppressed by the small R-parity
 violating coupling in addition to the Planck scale. (Notice that the
 R-parity violation in our scenario is even smaller than that
 considered in Ref.~\cite{Taka-Yama}. Thus, the lifetime of the
 gravitino in our case is much longer than that in their case.)  For
 such a long lifetime, the flux of the diffuse gamma ray generated by
 the gravitino decay is smaller than the observed
 value~\cite{Taka-Yama}.  Therefore, for $m_{3/2} > {\cal O}(1\kEV)$,
 the gravitino can be the dominant component of the dark matter in
 spite of the presence of the R-parity violation.\footnote{For
 $m_{3/2} < {\cal O}(1\kEV)$ the gravitino cannot be a cold dark
 matter, and hence the dominant component of the dark matter should be
 another particle or object.}

As for the next-to-lightest SUSY particle (NLSP), it can decay either
 (i) into gravitino via the usual R-parity conserving interaction, or
 (ii) into the standard model particles via R-parity violating
 couplings.  We have found that the partial decay rate of the latter
 one is much smaller than that of the former one, since the R-parity
 violations are extremely small.  (See Table.~\ref{table55} and
 \ref{table-1010}.)  Thus, the NLSP decay rate is determined by the
 former channel.  If the decay occurs during or after the big-bang
 nucleosynthesis (BBN), $t\simeq 1$--$100$ sec, it might spoil the
 success of the BBN~\cite{NLSPdecay}.  However, for $m_{3/2} < {\cal
 O}(1\MEV)$ the lifetime of the NLSP is shorter than 1 sec and this
 problem is avoided.

\section{Summary}
In this paper, we have studied nonanomalous discrete R-symmetry in GUT
without imposing the Giudice-Masiero condition. In the minimal SU(5)
GUT, $\mu\simeq 1\TEV$ is obtained only if the gravitino is ultralight
as $m_{3/2}\simeq 10^{-3}\EV$, and we find simple solutions $Z_{6R}$
and $Z_{30R}$. If a pair of ${\bf 5} \oplus {\bf 5^*}$ or ${\bf 10}
\oplus {\bf 10^*}$ are added to the minimal SU(5) GUT, we can find
many solutions predicting $m_{3/2} \gsim 10^{-3}\EV$.  Here, we
comment on the mass of these additional multiplets, $\xi + \bar{\xi}$,
which can be estimated since their charges are determined by the
anomaly cancellation condition.  [See Eqs.(\ref{EQ-xixi-55}) and
(\ref{EQ-1010-xixi}).]  The effective operator which induces the mass
of $\xi + \bar{\xi}$ is given by $W \simeq (\vev{W}/M_P^3)^{y'}M_P \xi
\bar{\xi}$.  We have checked that the mass $m_\xi\simeq
(m_{3/2}/M_P)^{y'}M_P$ is larger than the electroweak scale in all
cases we have discussed.

Since the fractional power of $\vev{W}$ is considered in this paper,
R-parity is necessarily violated, but R-parity violating couplings are
controlled by the symmetry.  In fact, the couplings were found to be
small enough to avoid the constraints from proton decay and neutrino
masses for some charge assignments.  Furthermore, it has also been
shown that dimension-five baryon- and lepton-number violating
operators are naturally suppressed.  Therefore, the proton stability
is ensured by the symmetry.

Low energy baryon- and/or lepton-number violating interactions might
cause a difficulty for baryogenesis since they might wash out the
baryon asymmetry together with the sphaleron~\cite{sphaleron}
process. However, we found that this is not the case for our scenario
because the R-parity violating couplings are so small that their
interactions have never been in thermal equilibrium.

The predicted gravitino masses were found to be in the range
$m_{3/2}\sim {\cal O}(10^{-3}\EV)$--${\cal O}(1\MEV)$ and the
gravitino is the LSP.  Since the lifetime of the gravitino is much
longer than the age of the universe, the gravitino can be the dominant
component of the dark matter.

As for the neutrino mass, there are two contributions in our
 framework, i.e., the seesaw mechanism and the R-parity violation.
 The R-parity violation can explain the neutrino mass scale if
 $\widehat{\mu}/\mu\sim {\cal O}(10^{-4}$--$10^{-7})$ and/or
 $\lambda_{\rm eff}\sim {\cal
 O}(10^{-4}$--$10^{-5})$~\cite{bilinear,trilinear,both}.  However, the
 predicted values of these couplings are either much larger or much
 smaller than these ranges.  The charge assignments predicting too
 large couplings are excluded.  Therefore, the neutrino mass scale
 should be generated by the seesaw mechanism, and contributions from
 the R-parity violation gives only tiny perturbation to it.

The discrete R-symmetry considered in this paper can explain the order
of magnitude of the $\mu$-term for light gravitinos. Hence, they can
be considered as solutions to the $\mu$-problem in low energy SUSY
breaking models such as the gauge mediation.

\subsection*{\it Note Added}
{\bf 1}~~ In our scenario, the predicted masses of the gravitino are
very small. We would like to stress that, if the gravitino is lighter
than about 100 keV, there is an interesting possibility to detect the
slow decay of the lightest neutralino into gravitino in future
collider~\cite{MakiOrito}.
\\
{\bf 2}~~ If the gravitino is lighter than the proton, proton can
decay into the gravitino via the R-parity violating coupling
$\lambda''U^c D^c D^c$. This leads to a stringent limit on the
coupling as $\lambda'' < 10^{-15}(m_{3/2}/{\rm
eV})$~\cite{Choi-Chun-Lee}. We found, however, that all the charge
assignments which satisfy the constraints discussed in the text also
satisfy this constraint, and hence the conclusion does not change. (We
are grateful to Ryuichiro Kitano for pointing out this constraint.)
\\
{\bf 3}~~
$B$-parameter in our case is at most of order loop suppression factor
times gaugino mass, which is induced at 2-loops in gauge
mediation. This implies that large $\tan\beta$ is preferred. (We wish
to thank Stephan Huber for useful comments and discussion.)

\section*{Acknowledgments}

The authors would like to thank T.~Yanagida for a collaboration at the
 early stage of this work and valuable discussions. We also wish to
 thank Rohini M. Godbole, Stephan Huber, and Ryuichiro Kitano for
 useful comments and discussion.  This work was supported by the Japan
 Society for the Promotion of Science (K.H. and N.M.).

\section*{Appendix}
In this appendix we briefly comment on the fractional power and
discrete symmetry. Although we have derived Eq.(\ref{mu1}) from
Eq.(\ref{m1}), it is nontrivial in the case of fractional power $y<1$
since the R-charges are defined under mod $N$ in the framework of
discrete $Z_{NR}$ symmetry.

Let us first discuss the vacuum expectation value of the
superpotential, $\vev{W}$. Suppose that a gauge singlet operator $X$
has a $Z_{NR}$ charge $2\al$ and it develops a vacuum expectation
value $\vev{X}\ll M_P$. Under the $Z_{NR}$ symmetry, it is also
possible to consider that $X$ has a charge $2\al + n N$ with $n \in
\z$. Then, in general, the superpotential can have the following
vacuum expectation value:
\bea
\vev{W}
 &=& 
\sum_{n> -2\alpha/N} c_n \left(\frac{\vev{X}}{M_P}\right)^{\frac{2\al}{2\al + nN}} M_P^3\,, 
\eea
where we have renormalized the mass dimension of $X$ operator to be
$1$. If all of the coefficients $c_n$ are of order one, the vacuum
expectation value of the superpotential becomes Planck scale, which is
inconsistent with almost vanishing cosmological constant or low energy
SUSY. Thus we expect that the right-hand side of the above equation is
dominated by a certain term with $n = n_0$:
\bea
\vev{W} &\simeq& c_{n_0} \left(\frac{\vev{X}}{M_P}\right)^{\frac{2\al}{2\al + n_0 N}} M_P^3\,.
\label{approx-W-vev}
\eea
The number $n_0$ is likely to depend on the operator $X$ as well as on
the origin of the $Z_{NR}$ symmetry, which we do not discuss in this
paper.  On the other hand, the operator relevant to the $\mu$-term is
also written in terms of $\vev{X}$:
\bea
W &=& 
\sum_{n> -2\alpha/N} c'_n 
\left(\frac{\vev{X}}{M_P}\right)^{\frac{2\al - h - \hb + r N}{2\al + nN}}
M_P H \bar{H}\,,
\eea
where $r$ is the minimum integer which gives $2\al - h - \hb + r N >
0$. Then, we naturally expect that the above operator is also
dominated by the term with $n = n_0$:
\bea
W &\simeq& c'_{n_0} 
\left(\frac{\vev{X}}{M_P}\right)^{\frac{2\al - h - \hb + r N}{2\al + n_0 N}}
M_P H \bar{H}\,.
\label{approx-mu-term}
\eea
Though it is possible that the fractional power which gives rise to
the $\mu$-term is different from the one responsible for the $\vev{W}$,
we argue that it is unnatural and assume that both of them are
dominated by the term with same $n = n_0$. Then, from
Eqs.(\ref{approx-W-vev}) and (\ref{approx-mu-term}), we obtain the
following expression of the $\mu$-term,
\bea
W &\simeq& \left(\frac{\vev{W}}{M_P^3}\right)^{\frac{2\al - h - \hb + r N}{2\al}}
M_P H \bar{H}
\,,
\eea
which leads to Eq.(\ref{m1}) with the $y$ given in (\ref{mu1}). The
same argument is also applied to the cases of baryon- and
lepton-number violating operators discussed in the text.

\end{document}